\documentclass[twocolumn,showpacs,preprintnumbers,prl]{revtex4}
\usepackage{graphicx}
\usepackage{dcolumn}
\usepackage{bm}
\usepackage{amsmath}
\usepackage{color}
\newcommand{\ymno}{YMn$_2$O$_5$}
\newcommand{\amno}{\emph{A}Mn$_2$O$_5$}

\begin{document}
\preprint{}
\title{Magnetic excitations in the low-temperature ferroelectric phase of  multiferroic \ymno~using inelastic neutron scattering}
\author{J.-H. Kim$^{1,2}$}
\author{M.A.van.der.Vegte$^3$}
\author{A. Scaramucci$^3$}
\author{S. Artyukhin$^3$}
\author{J.-H. Chung$^4$}
\author{S. Park$^{5,6}$}
\author{S-W. Cheong$^5$}
\author{M. Mostovoy$^3$}
\author{S.-H. Lee$^{1*}$}
\affiliation{$^1$Department of Physics, University of Virginia, Charlottesville,
VA 22904-4714, USA\\
$^2$Max-Planck-Institute for Solid State Research, Stuttgart, 70569, Germany\\
$^3$Zernike Institute for Advanced Materials, University of Groningen, AG Groningen, The Netherlands\\
$^4$Department of Physics, Korea University, Seoul 136-713, Korea\\
$^5$Rutgers Center for Emergent Materials and Department of Physics \& Astronomy, Rutgers University, Piscataway, NJ 08854, USA\\
$^6$Department of Physics, Chung-Ang University, Seoul 156-756, Korea}
\date{\today}

\begin{abstract}
 We studied magnetic excitations in a low-temperature ferroelectric phase of the multiferroic \ymno\ using inelastic neutron scattering (INS). We identify low-energy magnon modes and establish a  correspondence between the magnon peaks observed by INS and electromagnon peaks observed in optical absorption~\cite{SushkovPRL2007}. Furthermore, we explain the microscopic mechanism, which results in the lowest-energy electromagnon peak, by comparing the inelastic neutron spectral weight with the polarization in the commensurate ferroelectric phase.
\end{abstract}

\pacs{
78.70.Nx, 
75.30.Ds, 
75.85.+t, 
78.20.-e  
}
\maketitle

The coupling of spins with electric polarization in multiferroics gives rise to a wide variety of magnetoelectric effects actively studied over the past ten years~\cite{TKimura03,Hur04,Spaldin05,CheongNatMat2007,KhomskiiPhysics2009}. In many multiferroics, electric polarization is induced by magnetic structures formed when conventional magnetic orders are frustrated by lattice geometry and competition between spin interactions. Magnetic frustration can yield unusual excitations, such as spinons, monopoles and localized non-dispersive modes \cite{Morris2009,Fennell2009,Balents2010,LeeNature2002}. These magnetic excitations can couple to an electric field, which results in the electromagnon peaks in optical absorption~\cite{PimenovNaturePhys2006, SushkovPRL2007}.

Many spectacular magnetoelectric phenomena have been found in manganese oxides, such as mixing of magnons with acoustic phonons, clamping between ferroelectric and magnetic domain walls, and unusual magnetoelectric vortices in hexagonal YMnO$_3$~\cite{PetitPRL2007,FiebigNature2002,ChoiNatMat2010}. In orthorhombic \emph{A}MnO$_3$ compounds with \emph{A}=Tb, Dy, (Eu, Y), an incommensurate spiral spin order induces electric polarization through the relativistic `inverse Dzyaloshinskii-Moriya mechanism'~\cite{KenzelmannPRL2005,KatsuraPRL2005,SergienkoPRB2006,MostovoyPRL2006}. These compounds show magnetically-induced  polarization flops, giant magnetocapacitance and electromagnon peaks in
optical absorption~\cite{TKimura03,GotoPRL2004,PimenovNaturePhys2006,Aguilar09}. Smaller  cations on \emph{A}-sites, e.g., Ho, favor a commensurate collinear magnetic order  whose polarization, induced by the non-relativistic Heisenberg magnetoelectric coupling, is an order of magnitude larger than that of the spiral state~\cite{SergienkoPRL2006}.

A similar competition between collinear and non-collinear ferroelectric states is found in orthorhombic \amno~compounds. In this family, electric polarization is also larger in collinear states, while electromagnons and giant magnetocapacitance are observed in non-collinear ones~\cite{Hur04,HurPRL2004,SushkovPRL2007}. In \ymno, Mn spins order at 44 K in a periodically modulated structure with an incommensurate wave vector $(q_x,0,q_z)$. This state is paraelectric. Upon further cooling by a few degrees, first $q_z$ and then $q_x$ locks to a commensurate value and this commensurate (CM) phase, having the wave vector ${\bf q}_{\rm C} = (\frac{1}{2},0,\frac{1}{4})$, is ferroelectric~\cite{Kobayashi2004,INOM}. At 19 K the wave vector of the magnetic structure becomes again incommensurate: ${\bf q}_{\rm IC} = (0.48,0,0.288)$. This low-temperature incommensurate (LTI) state is also ferroelectric but has a much smaller polarization~\cite{Kobayashi2004,ChaponPRL2006}. This unusual sequence of transitions results from geometrical frustration in the Mn spin lattice~\cite{BlakePRB2005,HarrisPRL2008,Sushkov08}.

Magnetic orders in the CM and LTI phases were for some time a matter of controversy. However, several groups recently reported similar spin structures~\cite{Kobayashi2004,VecchiniPRB2008,Kim08,RadaelliPRB79}. In the CM phase spins are approximately collinear and lie nearly antiparallel along the $a$ axis, due to the relatively strong antiferromagnetic (AFM) exchange along this direction. The AFM chains are also discernable in the LTI state, which, however, is strongly non-collinear as the angles between spins in neighboring AFM chains are close to $\pm 90^{\circ}$. The LTI spin structure, which can roughly be characterized as a spiral where spins rotate in the $ab$ plane, is actually rather complex as it also contains considerable $bc$ and $ac$ spiral components~\cite{Kobayashi2004,Kim08,RadaelliPRB79}. A phenomenological analysis of the observed spin orders was used to show that the spontaneous electric polarization in the CM phase largely results from non-relativistic exchange interactions, while the much smaller polarization in the LTI phase has a sizable contribution from relativistic mechanisms~\cite{Kim08, RadaelliPRB79}. Furthermore, recent optical measurements revealed the presence of three electromagnon peaks in the optical absorption spectrum of the non-collinear LTI state, the nature of which was not fully understood \cite{SushkovPRL2007}.

In this Letter we report the results of inelastic neutron scattering (INS) for the LTI state of \ymno, providing the first information about magnetic excitations in \amno~of multiferroic materials. We also present a theoretical analysis of the observed low-energy magnon bands, identifying them as electromagnon modes. Good agreement with results of optical experiments \cite{SushkovPRL2007} clarifies the microscopic mechanism that couples spin waves to electric field in \amno~compounds.

\begin{figure}
\includegraphics[width=0.9\hsize]{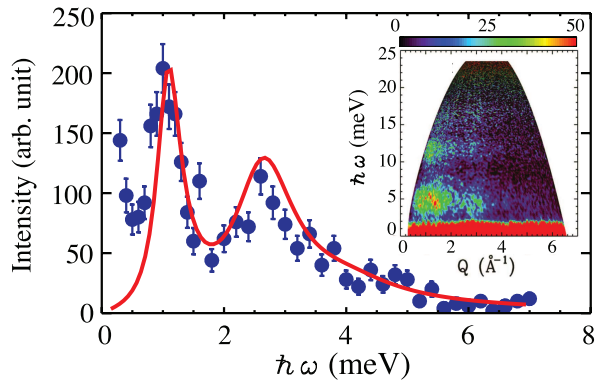}
\caption{Constant-$\bf{Q}$ scan on single crystals of \ymno~at $\mathbf{Q} =(1,0,0)+ \mathbf{q}_{\rm IC}$ at 4 K. Blue symbols are the neutron scattering data, while the  red line is the result of spin-wave calculations. The inset is a contour map obtained from a powder sample of \ymno~at 1.5 K.}
\label{constq}
\end{figure}

In order to study dynamic properties of \ymno, a 6 g powder sample and two single crystals with a total mass of about 1 g were used. The powder sample was sintered at 1180$^\circ$C for 60 h with intermediate grindings. X-ray powder diffraction results at room temperature (not shown) indicated single-phase \ymno~with a $Pbam$ orthorhombic unit cell~\cite{ChaponPRL2006}.  Single crystals were grown using B$_2$O$_3$-PbO-PbF$_2$~flux in a Pt crucible. The mixture was held at 1280$^\circ$C for 15 hours then cooled to 950$^\circ$C at 1$^\circ$C per hour. Both crystals used were characterized by neutron diffraction as described in Ref.~\cite{Kim08}. INS experiments on the powder sample were carried out in a liquid $^4$He cryostat using cold neutrons with the incident energy of 25.25 meV on the Disk-Chopper-Spectrometer (DCS) at the NIST Center for Neutron Research (NCNR). INS experiments on the single crystals were performed on the cold neutron triple-axis spectrometer, SPINS, and on the thermal neutron triple-axis spectrometer, BT7, at NCNR. The single crystals were co-aligned in the $(h,0,l)$ scattering plane within 1$^\circ$ and cooled in a 4 K closed-cycle refrigerator. For the INS measurements at SPINS, the energy of the scattered neutrons was set to $E_f$=5 meV, and a liquid nitrogen cooled Be filter was placed before the analyzer to eliminate higher order neutron contamination, while for the measurement at BT7, $E_f$=14.7 meV and a PG filter were used. 

\begin{figure}
\includegraphics[width=\hsize]{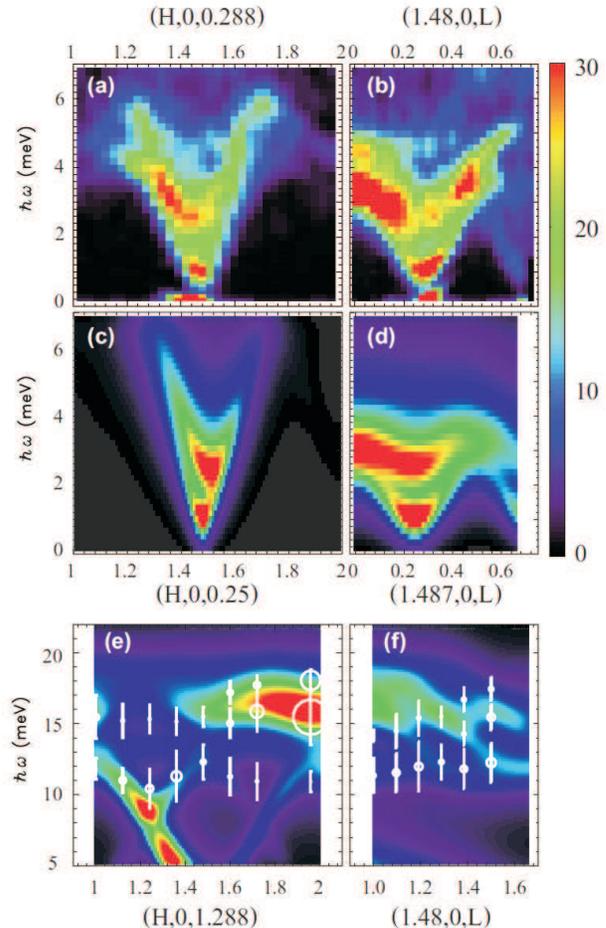}
\caption{Contour maps of the INS intensity, obtained at SPINS, along the (a) $H$- and (b) $L$-directions at 4 K. The intensities in (a) and (b) contours were multiplied by $\sqrt{\hbar \omega}$ to enhance the upper band. Corresponding calculated dispersions and intensities for parameters shown in Table~\ref{exc} convoluted with the instrument resolution ((c) and (d)). Experimental data taken at BT7 (open circles) and calculated intensities (contour map) for higher energy excitations from 10 to 20 meV along the (e) $H$- and (f) $L$-directions. The radiuses of the circles are proportional to the integrated intensities of the observed peaks.}
\label{ysingle_sw}
\end{figure}

The inset of FIG.~\ref{constq} shows a contour map of the INS intensity as a function of $Q$ and $\hbar\omega$, obtained from the powder sample at 1.5 K. In the LTI phase, there exist three broad bands of excitations: one is centered at $\sim$ 4 meV with a bandwidth of $\sim$ 5 meV, indicating that it contains more than one excitation mode, and the other two modes, which are weakly dispersive, are centered at $\sim$ 11 and 16 meV. FIG.~\ref{constq} shows typical energy $(\hbar\omega)$ scans at $\mathbf{Q} =(1,0,0)+ \mathbf{q}_{\rm IC}$ in the LTI phase. Up to $\hbar\omega = 7$ meV, there are at least three spin wave modes, resulting from a fit to centered at 0.98, 2.71, and 4.75 meV. We have performed similar $\hbar\omega$ scans at numerous wave vectors to map out the dispersions of the spin waves along two high symmetry directions, ($H$,0,0.288) and (1.48,0,$L$). The contour maps in FIG.~\ref{ysingle_sw} (a) and (b) summarize our results up to $\hbar\omega =$ 7 meV. Further constant-$\mathbf{Q}$ scans were performed at high energies around $\hbar\omega =$ 15 meV. The data were fit by Gaussians and the peak positions are plotted as open circles in FIG.~\ref{ysingle_sw} (e) and (f). As expected from the powder data shown in the inset of FIG.~\ref{constq}, for $\hbar\omega <$ 8 meV there are a few highly dispersive excitation modes, while for $\hbar\omega >$ 8 meV the excitations are weakly dispersive. In addition, there is a gap of $\sim$ 1 meV at the magnetic Bragg reflection due to magnetic anisotropy. The magnetic excitations are more dispersive along the $H$-direction than along the $L$-direction, which indicates that the magnetic interactions are stronger within the $ab$-plane than along the $c$-axis.

\begin{figure}
\includegraphics[width=\hsize]{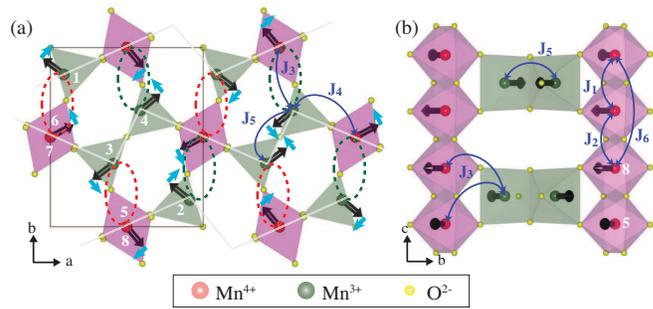}
\caption{Crystal and magnetic structures in the LTI phase of \ymno~projected onto the (a) $ab$- and (b) $bc$-plane. Thin white lines indicate the antiferromagnetic chains along the $a$ axis. Also shown are the exchange coupling constants $J_1 - J_6$ used to calculate the spin-wave spectrum. Black arrows indicate the spin directions for the numerically obtained ground state of the spin Hamiltonian Eq.(1). The light blue arrows in (a) represent the directions of the oscillating Mn spins in the optical phason mode. Green (red) ellipses show that the angle between neighboring spins is increased (decreased) by this mode.}
\label{js}
\end{figure}

Since the spin structure of the LTI phase is rather complex, we approximated the magnetic structure to a flat spiral in the $ab$ plane. We have numerically found the minimal-energy spin configuration in the magnetic unit cell corresponding to $ 39 \times 1 \times 4$ crystallographic unit cells (i.e., replacing ${\mathbf q}_{\rm IC} = (0.48,0,0.288)$ by  ${\mathbf q'}_{\rm IC} = (\frac{19}{39},0,\frac{1}{4})$) using the Heisenberg spin Hamiltonian for a spiral state  \begin{equation}
H=\sum_{\langle ij \rangle}J_{ij} \mathbf{S}_i \cdot \mathbf{S}_j
+\sum_{i} {\cal D}_i (S_i^c)^2,
\label{eq:Hamiltonian}
\end{equation}
where the sum is over pairs of Mn spins and $J_{ij}$ are the exchange coupling constants. We consider five nearest-neighbor interactions, $J_1$ to $J_5$ (defined as in Ref.~\onlinecite{ChaponPRL2006}) as well as the next-nearest-neighbor coupling along the $c$-axis, $J_{6}$~(see FIG.~\ref{js}). We also include the easy plane anisotropies,
${\cal D}_i = \Delta_1$~$(\Delta_2)$ on Mn$^{3+}$~(Mn$^{4+})$ sites. The optimal values of the coupling constants obtained by fitting the INS data are listed in Table~\ref{exc}. The red lines in FIG.~\ref{constq} and color maps of FIG.~\ref{ysingle_sw} (c)-(f) show the calculated INS spectrum. Despite the approximation made for the magnetic structure, our calculations reproduce the basic features of the observed magnetic excitations. The agreement between the calculations and the data is better for the $H$-dependence than for the $L$-dependence, probably due to the fact that ${\mathbf q'}_{\rm IC}$ and ${\mathbf q}_{\rm IC}$ are closer to each other in $H$~than $L$. 

\begin{table}[t]
\begin{tabular}{|c|c|c|c|c|c|c|c|c|c|}
  \hline
 &  J$_1$ & J$_2$ & J$_3$ & J$_4$ & J$_5$ & J$_6$ & $\Delta_{1}$ & 
$\Delta_{2}$\\
  \hline
  (meV) & -0.2& -1.5 & 0.15 & 3.75 & 2.75 &0.72 &0.13 & 0.11\\
  \hline
\end{tabular}
\caption{The parameters of the spin Hamiltonian Eq.(\ref{eq:Hamiltonian}).}
\label{exc}
\end{table}
The basic features of INS results with the wave vector $\mathbf{Q}$~produced by a spiral state can be understood by considering a circular spiral in the co-rotating-spin frame which has one coordinate axis parallel to the average spin vector and  another axis perpendicular to the spiral plane~\cite{Katsura07}. In this co-rotating spin-frame, the momentum transfer vector $\mathbf{Q}'$ is conserved. The magnetic excitations at $\mathbf{Q}'$ by the magnetic field {\em normal} to the spiral plane can be observed at $\mathbf{Q} =\mathbf{Q}'$ in the lab frame, while the same excitations by an {\em in-plane} magnetic field can be observed at $\mathbf{Q} = \mathbf{Q}' \pm \mathbf{q}_{\rm IC}$.

Using Eq.~(\ref{eq:Hamiltonian}) and the co-rotating-spin frame, we found that there are in total six modes that give rise to low-energy magnon bands in the LTI phase of \ymno. Three of them  -- a phason and two rotations of the spiral plane -- are generic for spiral magnets. The phason or sliding mode is a Goldstone mode of any incommensurate spiral state, having zero energy irrespective of the magnitude of higher harmonics in the spiral. It involves rotations of spins in the spiral plane and is excited either by an out-of-plane magnetic field with $\mathbf{Q} = 0$, or by an in-plane field with $\mathbf{Q} = \pm\mathbf{q}_{\rm IC}$. The other two low-energy spiral modes are rotations of the $ab$ spiral plane around the $a$ and $b$ axes, and have been found in other spiral magnets \cite{SenfPRL2007}. They can be excited either by the corresponding in-plane magnetic fields, $H_a$ and $H_b$,  with $\mathbf{Q} = 0$ (and, hence, can be observed as AFM resonances in optical experiments) or by an out-of-plane magnetic field with $\mathbf{Q} = \pm \mathbf{q}_{\rm IC}$. In the absence of an in-plane magnetic anisotropy,  these two modes are degenerate. The in-plane magnetic anisotropy induces higher spiral harmonics, lifting the degeneracy.  

These three modes have `optical' counterparts, which is a specific property of \amno~compounds due to the fact that spins in a crystallographic unit cell belong to two AFM chains along the $a$ axis. In the optical phason mode, spins in neighboring AFM chains rotate in opposite directions (see FIG.~\ref{js} (a)). The peculiar lattice  geometry of \amno~leads to an almost complete cancelation of exchange interactions between neighboring AFM chains~\cite{BlakePRB2005,Sushkov08}, resulting in near degeneracy of the `optical' and `acoustic' magnon branches.

\begin{figure}
\includegraphics[width=0.85\hsize]{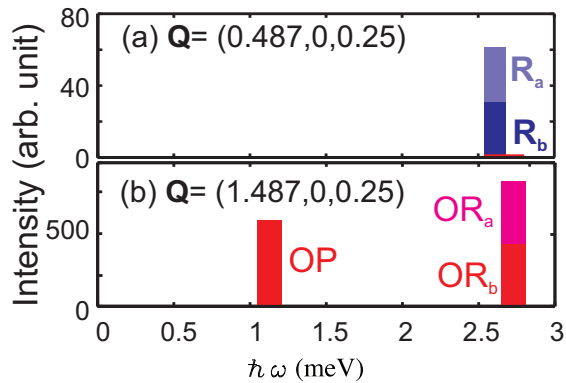}
\caption{Calculated INS intensities of the first three acoustic modes (blue bars): the phason (P), rotations around the $a$ and $b$ axes ($R_a$ and $R_b$), and their optical counterparts (O, red bars) at (a) $\mathbf{Q}=(\frac{19}{39},0,\frac{1}{4})$ and (b) $\mathbf{Q}=(1+\frac{19}{39},0,\frac{1}{4})$. Some modes which have almost zero intensities are not shown here.}
\label{Qcomp}
\end{figure}

FIG.~\ref{Qcomp} (a) shows the calculated INS energies and intensities of these six modes at $\mathbf{Q} = (\frac{19}{39},0,\frac{1}{4})$. At this $\mathbf{Q}$ position, the INS intensities corresponding to `acoustic' modes (blue) are large, while the intensities corresponding to the optical modes are hardly visible. At $\mathbf{Q} = (1+\frac{19}{39},0,\frac{1}{4})$ the situation is the opposite (see FIG.~\ref{Qcomp} (b)), which is related to the $\sim\frac{a}{2}$ shift between neighboring AFM chains. Comparing these results with the experimental data at  $\mathbf{Q} = (1.48,0,0.288)$, we have identified the $\hbar \omega \sim 1$meV peak as the optical phason which corresponds to the peak at $7.2$~cm$^{-1}$ observed in optical absorption~\cite{SushkovPRL2007}. The two optical rotation modes account for the broad peak around 2.5 meV in our INS data. This interpretation is supported by the observation of two close peaks at 16 cm$^{-1}$ and 21 cm$^{-1}$ in optical experiments~\cite{SushkovPRL2007}. These three peaks observed in the optical experiment are electromagnons as they are excited by electric field of light, $E \| b$ \cite{SushkovPRL2007}. It was previously speculated that the Heisenberg exchange mechanism that induces the electric polarization $P \| b$ in the CM phase also couples the optical phason mode to $E \|b$~\cite{Sushkov08, RadaelliPRL2008}. The corresponding invariant for the $Pbam$ space group has the form
\begin{align}
H_{\rm ME}= -gE_b & \sum_{n} \left[(\mathbf{S}_{1,n}-\mathbf{S}_{2,n-a}) \cdot (\mathbf{S}_{6,n}+\mathbf{S}_{7,n}) \right.  \nonumber \\
& + \left.(\mathbf{S}_{3,n}-\mathbf{S}_{4,n-b}) \cdot (\mathbf{S}_{5,n}+\mathbf{S}_{8,n}) \right\},
\label{eq:magnetoelec}
\end{align}
where the indices 1-8 label Mn ions in the unit cell (see FIG.~\ref{js}), while $n$ labels unit cells. The electric field of light, $E \|b$, modulates interchain interactions along the $b$-direction, e.g., $J_{3,8} \rightarrow J_{3,8} + \Delta J$, while $J_{4,8+b} \rightarrow J_{4,8+b} - \Delta J$, which induces the relative rotation of spins in neighboring AFM chains (see FIG.~\ref{js}). To prove this scenario we find the coupling constant $g$ using the value of polarization in the CM phase, $P_b = 1000$ $\mu$C$\cdot$~m$^{-2}$, and calculate the spectral weight, $S$, of the optical phason. This gives $S =194$~cm$^{-2}$, which is close to the experimental value of $170$~cm$^{-2}$~\cite{Sushkov08}. For large deviations of the LTI spin structure from the flat spiral,  Eq.(\ref{eq:magnetoelec}) can also explain the excitation of the spiral rotation modes.

In summary, we identified six low-energy magnon modes and determined the exchange interactions in \ymno. Three of the six magnon modes are common to all spiral magnets, while the remaining three result from magnetic frustration, which is the source of the complex magnetic behavior and ferroelectricity found in \amno~compounds. The latter three modes were identified with the electromagnons observed in optical conductivity measurements.

The works at UVA and NCNR were supported by the US NSF under Agreement No. DMR-0903977 and DMR-0454672, respectively. The work at Rutgers was supported by the DOE Grants No. DE-FG02-07ER46382. JHC is supported by National Research Foundation of Korea (Grant No. 2010-0018369).

* Email: shlee@virginia.edu

\end{document}